\documentclass[prl,superscriptaddress,floatfix,letterpaper,twocolumn,aps,showpacs]{revtex4-1}
\usepackage[utf8]{inputenc}
\usepackage{natbib}
\usepackage{amsmath}
\usepackage{amsbsy}
\usepackage{amssymb}
\usepackage{scrextend}
\usepackage{graphicx}
\usepackage{color}
\usepackage{xcolor}

\usepackage[T1]{fontenc}

\newcommand{\BNOO}{$\mathrm{Ba}_{2}\mathrm{Na}\mathrm{Os}\mathrm{O}_{6}$}
\begin{document}
%
%
\title{Phase transition preceding magnetic long-range
order in the double perovskite $\mathrm{Ba}_{2}\mathrm{Na}\mathrm{Os}\mathrm{O}_{6}$}

%
\author{Kristin Willa}
\affiliation{Materials Science Division, Argonne National Laboratory, 9700 South Cass Avenue, IL 60439, USA}
\affiliation{Institute for Solid State Physics, Karlsruhe Institute of Technology, Karlsruhe D-76021, Germany}
\author{Roland Willa}
\affiliation{Materials Science Division, Argonne National Laboratory, 9700 South Cass Avenue, IL 60439, USA}
\affiliation{Institute for Theoretical Condensed Matter physics, Karlsruhe Institute of Technology, Karlsruhe D-76131, Germany}
\author{Ulrich Welp}
\affiliation{Materials Science Division, Argonne National Laboratory, 9700 South Cass Avenue, IL 60439, USA}
\author{Ian R. Fisher}
\affiliation{Department of Applied Physics and Geballe Laboratory for Advanced Materials, Stanford University, California 94305, USA}
\affiliation{Stanford Institute for Materials and Energy Sciences, SLAC National Accelerator Laboratory, 2575 Sand Hill Road, Menlo Park, California 94025, USA}
\author{Andreas Rydh}
\affiliation{Department of Physics, Stockholm University, SE-106 91 Stockholm, Sweden}
\author{Wai-Kwong Kwok}
\affiliation{Materials Science Division, Argonne National Laboratory, 9700 South Cass Avenue, IL 60439, USA}
\author{Zahir Islam}
\affiliation{Advanced Photon Source, Argonne National Laboratory, 9700 South Cass Avenue, Il 60439, USA}
\date{\today}

\begin{abstract}
Recent theoretical studies [Chen \emph{et al.}, Phys. Rev. B 82,
174440 (2010), Ishizuka \emph{et al.}, Phys. Rev. B 90, 184422
(2014)] for the magnetic Mott insulator $\mathrm{Ba}_{2}\mathrm{Na}\mathrm{Os}\mathrm{O}_{6}$ have proposed a low-temperature order parameter that breaks lattice rotational symmetry without breaking time reversal symmetry leading to a nematic phase just above magnetic ordering temperature. We present high-resolution calorimetric and magnetization data of the same $\mathrm{Ba}_{2}\mathrm{Na}\mathrm{Os}\mathrm{O}_{6}$ single crystal and show evidence for a weakly field-dependent phase transition occurring at a temperature of $T_s \approx 9.5 $K, above the magnetic ordering temperature of $T_c \approx 7.5 $K. This transition appears as a broadened step in the low-field temperature dependence of the specific heat. The evolution of the phase boundary with applied magnetic field suggests that this phase coincides with the phase of broken local point symmetry seen in high field NMR experiments [Lu \emph{et al.}, Nat.\ Comm.\ \textbf{8} 14407 (2017)]. Furthermore, the magnetic field dependence of the specific heat provides clear indications for magnetic correlations persisting at temperatures between $T_c$ and $T_s$ where long-range magnetic order is absent giving support for the existence of the proposed nematic phase.
\end{abstract}

\maketitle
The simultaneous occurrence of strong spin-orbit coupling (SOC), orbital degrees of freedom and electronic correlations gives rise to a remarkable variety of electronic states \cite{Witczak-Krempa2014, Rau2016} ranging from unconventional superconducting and magnetic phases to spin liquids and topological states of matter. This richness is at display in the rock-salt ordered double perovskites with general composition A$_2BB^{’}$O$_6$ \cite{Vasala2015, Marjerrison2016} where B denotes a non-magnetic ion and B$^{’}$ an ion with partially filled 4d or 5d orbitals. The double perovskite structure can be viewed as two interpenetrating fcc-lattices of O-octahedra centered around the B and B$^{’}$ ions, respectively, resulting in a geometrically frustrated arrangement of magnetic moments with relatively large separation. Consequently, magnetic exchange energy, spin-orbit coupling and correlation energies are of comparable magnitude such that a diverse range of ground states can arise. In particular, the strong SOC induces highly anisotropic non-Heisenberg exchange interactions \cite{Kugel1982} which depend on the orbital occupancy and are therefore strongly tied to orbital order \cite{Xiang2007, Gangopadhyay2015, Chen2010, Svoboda2017, Xu2016, Ishizuka2014, Romhanyi2017, Ahn2017, Iwahara2018, Dodds2011}. 

Among these double perovskites, the Mott-insulating {\BNOO} has garnered particular interest as it undergoes a ferromagnetic transition below 8 K \cite{Stitzer2002, Erickson2007, Steele2011} with a (110) easy axis \cite{Erickson2007} that is not expected in standard Landau theory of magnetic anisotropy in a cubic system \cite{Chen2010} and a negative Curie-Weiss temperature indicative of antiferromagnetism \cite{Erickson2007}. The formal valence of Os is $7^{+}$ corresponding to a 5d$^1$ state although hybridization between Os-5$d$ and O-$2p$ states has been recognized as an important factor determining the magnetic state of Os \cite{Xu2016, Ahn2017} and may be responsible for the small saturation moment of less than $0.2\mu_{B}$/Os. The 5$d^1$ configuration in an octahedral crystal field in the presence of strong SOC leads to a four-fold degenerate $t_{2g}$ ground state corresponding to an effective angular momentum of 3/2. This state is unstable against a Jahn-Teller distortion \cite{Ishizuka2014, Xu2016, Dodds2011} that induces a tetragonal paramagnetic phase with a two fold degenerate ground state as also indicated by measurements of the entropy balance across the magnetic transition yielding a value close to $R \ln 2$ rather than $R \ln 4$ \cite{Erickson2007}. Furthermore, recent high-field NMR experiments \cite{Lu2017, Liu2018, Liu2018b} reveal that the magnetically ordered state of {\BNOO} is composed of an exotic non-collinear magnetic structure in which successive ferromagnetic Os-Na layers that subtend a canting angle of $\approx 134^{\circ}$ resulting in a small net magnetic moment along (110). This long-range magnetic order is driven by orbital ordering and has an over-all orthorhombic symmetry. In addition, at high magnetic fields and several kelvin above the magnetic transition temperature, signatures of orbital ordering into a state that preserves time reversal symmetry but breaks the four-fold lattice symmetry of the high-temperature paramagnetic phase have been observed in NMR experiments \cite{Lu2017, Liu2018, Liu2018b}. However, the evolution of this phase at low fields has not been reported so far.


Here, we present high-resolution calorimetric and magnetization data obtained on the same {\BNOO} single crystal. Using general thermodynamic considerations and the Fisher formula \cite{Fisher1962} we separate magnetic and non-magnetic contributions to the specific heat, and present evidence for a largely field-independent phase transition occurring at $T_s \approx 9.5 $K, slightly above the magnetic ordering temperature of $T_c \approx 7.5$ K. This transition appears as a broadened step in the temperature dependence of the specific heat, and its evolution with applied magnetic field suggests that this phase corresponds to the broken local point symmetry phase seen in NMR experiments in high fields \cite{Lu2017, Liu2018, Liu2018b}. Our specific heat results, in conjunction with the NMR findings, indicate that this new phase between $T_c$ and $T_s$ is nematic in origin as predicted by theoretical models \cite{Chen2010, Ishizuka2014}. Here we establish its low field phase boundary. The magnetic field dependence of the specific heat reveals that magnetic correlations persist into the nematic phase even though long-range magnetic order is absent.

\begin{figure}[tb]
\includegraphics[width=0.92\linewidth]{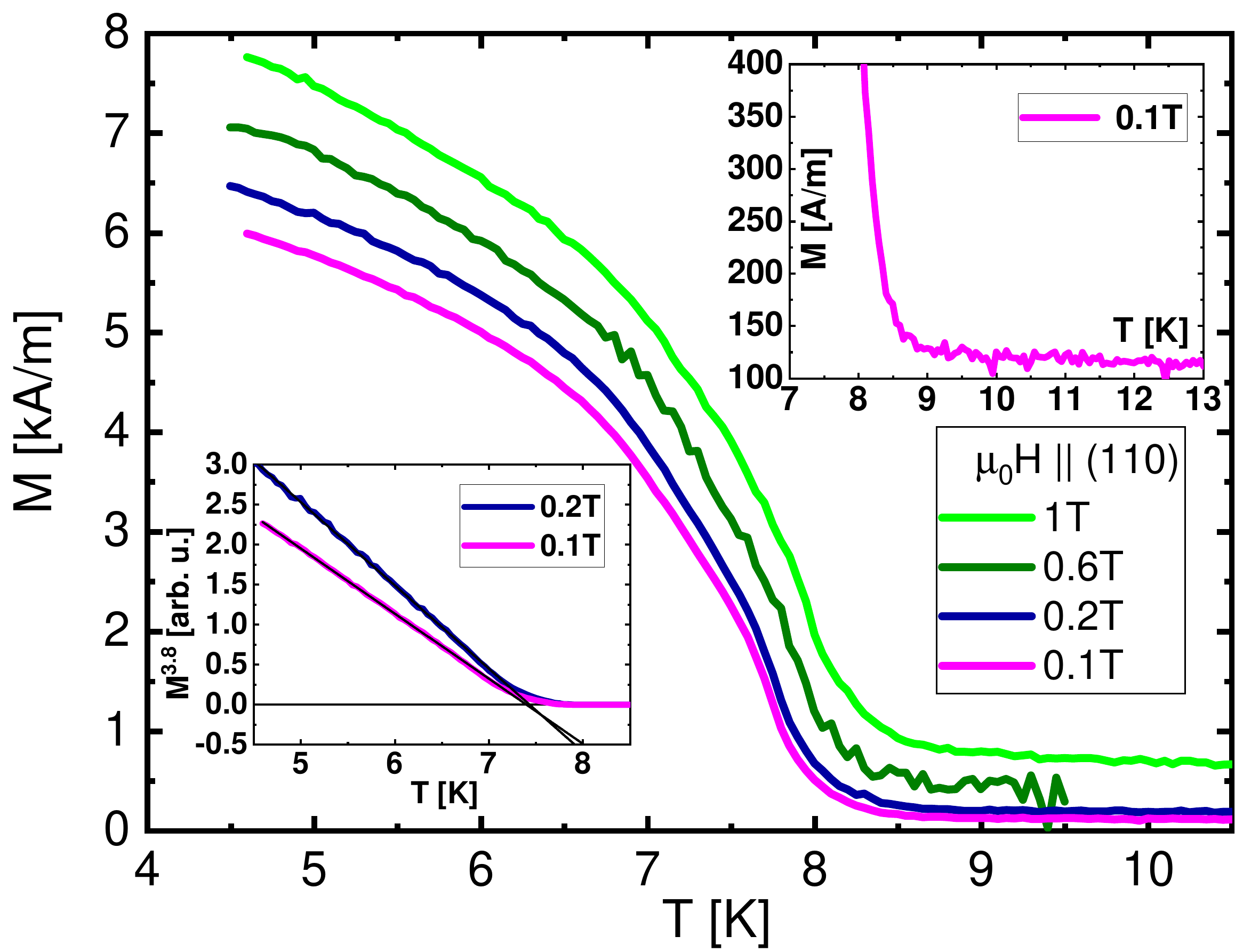}
\caption{
Temperature-dependent magnetization for a {\BNOO} single crystal in various fields applied along (110). Approaching the transition temperature from below, the magnetization vanishes as $M \propto (1 - T/T_{c})^{\beta}$, with $\beta \approx 0.26$; different from a mean-field value $\beta_{\mathrm{mf}} = 1/2$ (bottom inset). The top inset shows the magnetization on largely expanded scales and highlights that the tail of the magnetic transition extends to beyond 9K.
}
\label{fig:mag}
\end{figure}

\begin{figure}[h!]
\includegraphics[width=0.92\linewidth]{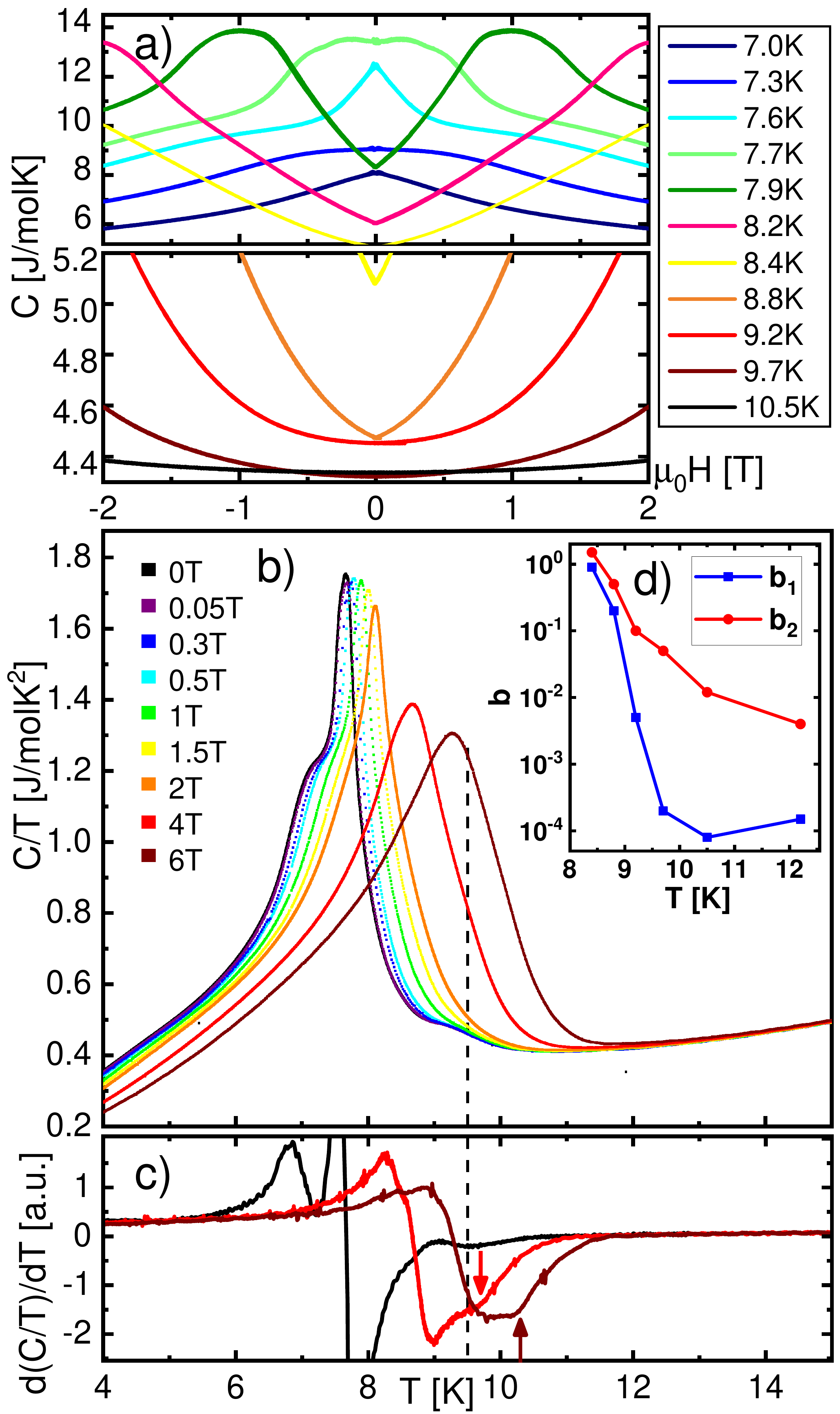}
\caption{\textbf{a)} Isothermal magnetic field dependence of the specific heat of {\BNOO} single crystal. \textbf{b)} Temperature dependence of the specific heat of {\BNOO} single crystal at fixed magnetic fields along the (110) direction. The main peak corresponds to the magnetic phase transition, while we attribute high-temperature shoulder near $9.5\mathrm{K}$ (indicated by the dotted line) to a nematic transition. No field dependence is observed for the nematic transition up to 0.5T. For higher fields, this transition is masked by the main magnetic transition. \textbf{c)} Temperature derivative of the specific heat data at 0T, 4T, and 6T. The shoulder-feature in $C/T$ corresponds to a negative-going dip in the derivative as seen in the 0-T trace at the location of the dashed line. In fields of 4 T and 6 T this dip has shifted to 9.7 K and 10.3 K, respectively, as marked by the arrows revealing the nematic transition. \textbf{d)} Temperature dependence of the parameters $b_1$ and $b_2$ entering the fit $C(H,T) = C_0(T) + b_1(T)\mu_{0}|H| + b_2(T)(\mu_{0}H)^2 + b_4(T)(\mu_{0}H)^4$.}
\label{fig:spec-heat}
\end{figure}

Single crystals of {\BNOO} were grown by molten hydroxide flux, following the method presented in Ref.\ \cite{Stitzer2002}. We polished a plate-like sample perpendicular to the (001) direction down to approximately 34 $\mu $m thickness. The crystal was then successively attached with its polished surface onto a quartz fiber for Superconducting Quantum Interference Device (SQUID) magnetometry and onto a nanocalorimetric membrane for specific heat measurements. In both setups, the external magnetic field was applied along the crystallographic (110) direction, the magnetic easy axis \cite{Erickson2007}. The magnetization data reveals the onset of magnetic order at temperatures below 8 K as evidenced by the steep increase in the magnetisation (see Fig.\ \ref{fig:mag}), in agreement with earlier reports \cite{Stitzer2002, Erickson2007, Steele2011}. Over a remarkably wide temperature range, the magnetisation follows the relation $M(T) \approx (1-T/T_c)^{\beta}$ with $\beta \approx 0.26$, see Fig.\ \ref{fig:mag}. Determining the magnetic transition temperature $T_c = 7.5 $K from this scaling is in good agreement with the calorimetric results presented below.

\begin{figure}[tb]
\includegraphics[width=0.92\linewidth]{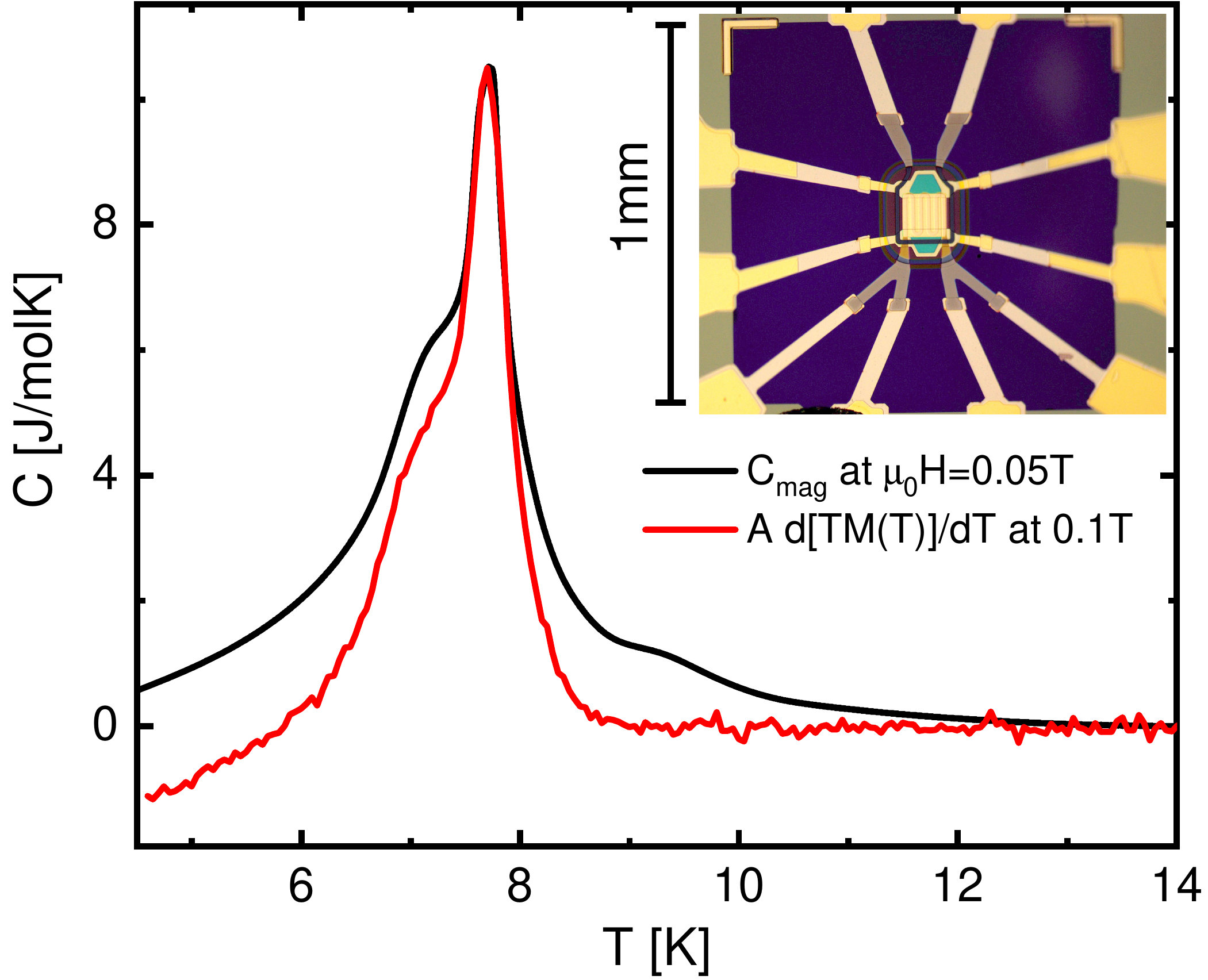}
\caption{
Comparison of $A \partial [T M(T)]/\partial T$  evaluated in a field of 0.1 T (red) and the measured specific heat after subtracting a phonon background (black). A moving average of five points has been applied to the magnetization data prior to taken the derivative. Here, the coefficient A has been adjusted to match the peak heights. The location and shape of the main magnetic peak and the low-temperature shoulder are well reproduced. However, the high-temperature shoulder near 9.5 K is absent in the data derived from the magnetization suggesting that this feature is largely non-magnetic in origin. The nanocalorimeter platform is shown in the inset.
}
\label{fig:fisher}
\end{figure}

Specific heat data were obtained using a membrane-based ac-calorimeter as described in Refs.\ \cite{Willa2017, Tagliati2012} and pictured in the inset of Fig.\ \ref{fig:fisher}. Through the membrane's ac heater, the sample is subjected to a modulated heat signal, and the resulting temperature oscillation is recorded as a second harmonic signal. Along a temperature scan, the oscillation frequency is kept constant (around 1 Hz) and the heater power is adjusted to keep a constant relation between absolute temperature and temperature modulation amplitude. The specific heat data in zero applied field and in fields up to 6 T is shown in Fig.\ \ref{fig:spec-heat}b). In zero field, we observe a sharp central peak at 7.5 K that signals the transition into the magnetically ordered state which is accompanied by a small shoulder on the high temperature side, consistent with an earlier zero field specific heat study \cite{Erickson2007}. Additionally a low-temperature shoulder can be seen which, for increasing fields, moves, together with the main peak, to higher temperatures (at a rate of 0.26 K/T) until they merge in high fields. This is indicative for both features being magnetic in origin, albeit the causes of the shoulder are not known at present. On the contrary, the high-temperature shoulder does not depend on field strength for small fields. By viewing this shoulder as a broadened step centered near 9.5 K we estimate the associated step height of 50 mJ/molK$^2$. At fields higher than 1.5 T, this shoulder is masked by the main peak moving to higher temperatures. However, as revealed by the temperature derivative of the 4T and 6T data, signatures of the shoulder are still discernible in high applied fields residing on the high-temperature slope of the main peak, as marked by the arrows in Fig.\ \ref{fig:spec-heat}c). At 4T, the sharp dip at 9 K originates from the main inflection point of the $C/T$-curve, while the feature due to the shoulder occurs near 9.7 K (see arrow). In 6T, the inflection point as shifted to $\approx$ 9.8 K and the shoulder to 10.3 K. For reference, the 0T data has been included as well.

In order to gain further insight into the origin of the calorimetric response we apply the Fisher relation \cite{Fisher1962}
\begin{align}\label{eq:Fisher}
   C(T) \simeq A\, \{ \partial [\,T \chi(T)\,] / \partial T \}
\end{align}
which links the magnetic specific heat $C_m$ associated with an antiferromagnetic transition to the temperature dependence of the zero-field susceptibility $\chi = \partial M/ \partial H)|_{H=0}$. Here, the parameter $A$ depends only weakly on temperature. This relation has been applied extensively to the investigation of antiferromagnetic transitions, for instance in dysprosium aluminum garnet \cite{Wolf1964}, HoAsO$_4$ and GdVO$_4$ \cite{Becher1975, Ni2008}, and more recently in Fe-based superconductors \cite{Ni2008, Chu2009, Rotundu2011}. The work on the Fe-based superconductors, in particular, revealed that Eq.\ \eqref{eq:Fisher} yields a good description of magnetization and caloric data also at temperatures beyond the vicinity of the antiferromagnetic transition. Nevertheless, its applicability to {\BNOO} may be questioned since this material is not a simple antiferromagnet but rather dominated by the ferromagnetic component. However, as described above, the large canting angle between ferromagnetic layers induces pre-dominantly antiferromagnetic correlations as evidenced for instance by the negative Curie Weiss temperature [$\Theta \!=\! -13$K for $H||$(110)] and by a comparatively small ferromagnetic moment of less than $0.2 \mu_B /$Os \cite{Erickson2007}. We therefore contend that relation \eqref{eq:Fisher} is suitable for analyzing the magnetic and caloric properties of this material. We apply Eq.\ \eqref{eq:Fisher} to the magnetization data taken in a field of 0.1T and approximate the susceptibility as $\chi = M/H$. From the measured specific heat we subtracted a background ($C_{backgd} = \alpha' T + \alpha T^3$). As shown in Fig.\ \ref{fig:fisher}, the peak position and shape as well as the low-temperature shoulder are well reproduced. However, the high-temperature shoulder is absent in the specific heat curve derived from the magnetization data suggesting that this feature is largely non-magnetic, possibly structural, in origin and does not change the magnetic level degeneracy. This in combination with the results from Ref. \cite{Erickson2007} points to a scenario of a transition at $T_s$ between non-cubic states. Shoulders in the specific heat preceding a magnetic transition similar to those observed here have been reported in studies of the phase transitions of Ba(Fe$_{1-x}$Co$_x$)$_2$As$_2$ where they were attributed to a second order structural transition between the high-temperature tetragonal paramagnetic phase and an orthorhombic nematic phase \cite{Chu2009, Rotundu2011}. The magnetization of Ba(Fe$_{1-x}$Co$_x$)$_2$As$_2$ does display a change in slope at the structural transition such that the application of Eq.\ \eqref{eq:Fisher} yields the shoulder in the specific heat, in contrast to the data presented here on {\BNOO}. This behavior of Ba(Fe$_{1-x}$Co$_x$)$_2$As$_2$ has been ascribed to the pronounced magneto-elastic coupling reported in Fe-based superconductors \cite{Kuo2012, Fernandes2014, Fernandes2013}. Albeit strong magneto-elastic coupling is expected for {\BNOO} \cite{Ishizuka2014, Xu2016, Iwahara2018} as described above, our results imply that this coupling is nevertheless weaker than in Fe-based superconductors.
\begin{figure}[t!]
\includegraphics[width=0.92\linewidth]{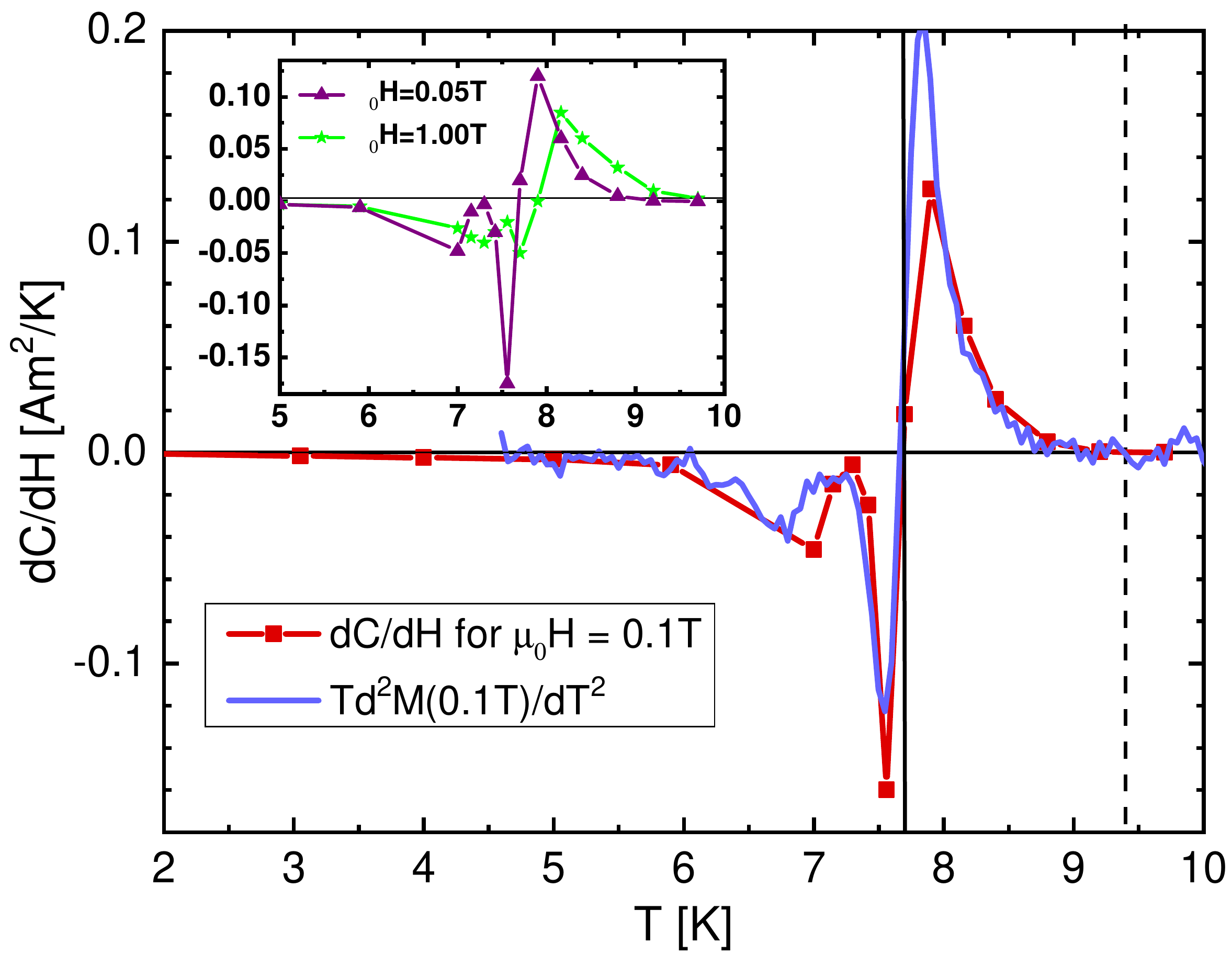}
\caption{
The thermodynamic relation \eqref{eq:maxwell} is tested here by plotting the derivatives $\partial C / \partial H$ (red) and $T \partial^{2}M/\partial T^{2}$ (blue) against temperature and for a fixed field of 0.1T. The absence of a feature above 7.5K is evidence for the structural origin of the high-temperature shoulder observed in the specific heat, see Fig.\ \ref{fig:spec-heat}. The inset shows $\partial C / \partial H$ for two other fields 0.05T and 1T allowing us to track the field dependence of the low temperature shoulder and peak position (zero crossing) of the specific heat.
}
\label{fig:derivatives}
\end{figure}

The good agreement between specific heat measured directly and that derived from magnetization data using the Fisher relation \eqref{eq:Fisher} allowed to qualitatively disentangle magnetic from non-magnetic calorimetric contributions. We reach beyond this approach by performing detailed measurements of the field-dependence of the specific heat and analyzing the data according to the thermodynamic relation \cite{Stanley1971},

\begin{figure}[t!]
\includegraphics[width=0.92\linewidth]{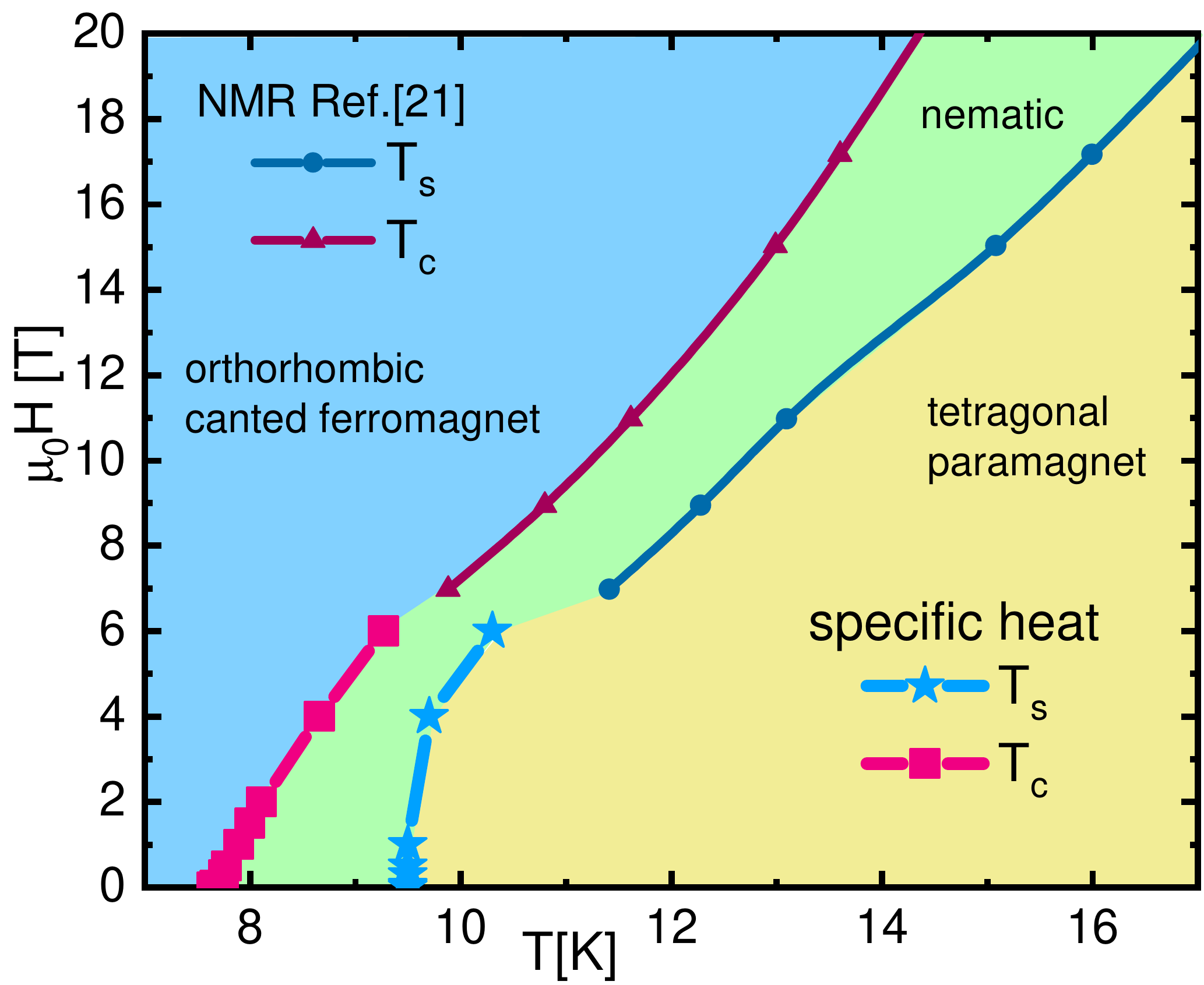}
\caption{
Magnetic phase diagram of {\BNOO} as deduced from specific heat measurements containing the orthorhombic canted ferromagnetic phase at low temperatures, the nematic (i.e., orthorhombic paramagnetic) phase at intermediate temperatures, and the tetragonal paramagnetic phase at high temperatures. Shown are the fields of the peak in the specific heat marking the transition to long-range magnetic order (magenta) and the high-temperature shoulder (light blue) marking the beginning of the nematic phase. Also included are the high-field data from NMR experiments \cite{Liu2018b} for the magnetic transition (purple) and nematic phase (phase of broken local point symmetry) (dark blue). NMR data were taken in fields applied along a (001) direction while the specific heat was measured in (110)-fields.
}
\label{fig:phasediag}
\end{figure}
\begin{align}\label{eq:maxwell}
   \partial C / \partial B &= T [\partial^{2} M / \partial T^{2}],
\end{align}
derived from the Gibbs free energy $G(T,H) = U - S T + p V - M H$. The upper panel of figure \ref{fig:spec-heat} shows the field dependence of the specific heat measured at temperatures near the zero-field magnetic transition. Its non-monotonic field dependence clearly reflects the shift of the magnetic transition to higher temperatures in increasing field. Above the magnetic transition and in the field range of $|\mu_{0} H| < 1$T the specific heat can be well parametrized by $C(H,T) = C_0(T) + b_1(T)\mu_{0}|H| + b_2(T)(\mu_{0}H)^2 + b_4(T)(\mu_{0}H)^4$. The $|H|$-term accounts for the cusp-like variation seen at temperature below 9.5 K. The (in leading order) quadratic dependence in the high-temperature phase implies that in the limit of small fields the specific heat is indeed field independent. More quantitatively, the rapid disappearance of the linear coefficient $b_1$ at 9.5 K as compared to the quadratic coefficient $b_2$, can be seen in Fig.\ \ref{fig:spec-heat}d). While $b_1$ falls to zero within our experimental resolution $\sim 10^{-4}$,  $b_4$ drops from $b_4 \approx 0.05$ at 9.2 K to values of $4\times 10^{-4}$ at 10.5 K as well, leaving the quadratic coefficient $b_2$ as the leading order term. A quadratic field-dependence of the specific heat, such as observed above 9.5K, is expected for a system with field-independent susceptibility, $M(T,H) = \chi(T)H$. In this case, Eq.\ \ref{eq:maxwell} can be integrated to yield $C(H,T) = C_0(T) + (1/2) \mu_0 T H^2 \partial^2\chi(T)/\partial T^2$. Thus, the salient feature of the data in Fig.\ \ref{fig:spec-heat}d) is that at low applied fields a non-linear susceptibility and magnetic correlations emerge when cooling below $9.5$ K. Within our resolution, these features remain unresolved in the magnetization, inset in Fig.\ \ref{fig:mag}. Compared with a simple Curie-Weiss law, the quadratic coefficient, $b_2$, in the specific heat drops much faster which is consistent with strong deviations of the susceptibility from a Curie-Weiss behavior below 20 K reported elsewhere \cite{Erickson2007}.

Eq.\ \eqref{eq:maxwell} can be used to validate the experimental data. As shown in Fig.\ \ref{fig:derivatives}, the second derivative of the magnetization (obtained from Fig.\ \ref{fig:mag}) is in excellent quantitative agreement with the first derivative of the specific heat (data from Fig.\ \ref{fig:spec-heat}). The calibration-free agreement underlines the consistency across the different measurements while the absence of any features above 9.5K (marked by the dotted line) confirms again the non-magnetic origin of the high temperature shoulder.

We use the specific heat data from Figs.\ \ref{fig:spec-heat} and \ref{fig:derivatives} to construct the phase diagram as displayed in Fig.\ \ref{fig:phasediag}. Shown are the fields of the peak in the specific heat (respectively the zero crossing in $\partial C/\partial H$) marking the transition to long-range magnetic order and the high-temperature shoulder. Also included are the high-field data from NMR experiments for the magnetic transition and the transition to a phase of broken local point symmetry \cite{Liu2018b}. The evolution of the latter boundary with applied magnetic field suggests that the high-temperature shoulder represents the continuation of the broken local point symmetry phase to low fields. As this transition involves the distortion of oxygen-octahedra and is thus primarily structural in origin, an almost field-independent transition may be expected at low fields. At the same time, the entanglement of spin, orbital and lattice degrees of freedom inherent to the Mott-insulating double perovskites may account for the emergence of correlation effects and the field dependence of the specific heat in particular in fields above 4 T. Indeed, recent ab-initio calculations \cite{Iwahara2018} indicate that due to this entanglement the lattice vibrations and atomic positions become dependent on an effective magnetic field which contains contributions from an exchange field arising from magnetic correlations and from the applied field. The shape of the phase boundary suggests that this exchange corresponds to an internal field scale of approximately 4 T as in applied fields higher than this cross over field the phase boundary is clearly field dependent.

In conclusion, we have presented high-resolution calorimetric and magnetization data obtained on the same {\BNOO} single crystal. By differentiating magnetic and non-magnetic contributions we find at low fields a largely field-independent phase transition occuring a few kelvin above the magnetic transition. At higher fields this transition becomes field dependent and the thus obtained phase boundary  agrees well with a recent high field NMR study that reports a phase of broken local point symmetry above the magnetic ordering temperature. Our results imply that the transition into the nematic state is indeed a thermodynamic phase transition. Furthermore, the magnetic field dependence of the specific heat gives clear indication for magnetic correlations persisting in the nematic phase $T_s>T>T_c$ even though long-range magnetic order is absent.
\begin{acknowledgments}
We thank Han-Oh Lee for help with crystal growth and Paul Wiecki as well as Frederic Hardy and Christoph Meingast for useful discussions.
This work was supported by the U.S.\ Department of Energy, Office of Science, Basic Energy Sciences, Materials Sciences and Engineering Division. Both K.\ W. and R.\ W. acknowledge support from the Swiss National Science Foundation through the Early Postdoc.Mobility program. Work at Stanford University was supported by the DOE, Office of Basic Energy Sciences, under Contract No. DE-AC02-76SF00515. This research used resources of the Advanced Photon Source, a U.S. Department of Energy (DOE) Office of Science User Facility operated for the DOE Office of Science by Argonne National Laboratory under Contract No. DE-AC02-06CH11357.
\end{acknowledgments}

\onecolumngrid
\vspace{2em}
\twocolumngrid

\bibliography{KristinBibliography}

\end{document}